\documentclass{article}
\usepackage[utf8]{inputenc}
\usepackage{amsmath,amssymb}
\usepackage{mathtools}
\usepackage{multirow}
\usepackage[hidelinks]{hyperref}
\usepackage{qcircuit}
\usepackage{xargs}
\usepackage{tikz}
\usepackage{todonotes}
\usepackage{comment}
\usepackage{authblk}
\usetikzlibrary{calc,positioning}
\usetikzlibrary{arrows,decorations.markings}

\newcommand{\ket}[1]{| #1 \rangle}

\newcommand{\ketbra}[2]{| #1 \rangle \langle #2 |}

\newcommand{\kron }{\otimes}

\newcommand{\Hl}{\mathcal{H}}
\newcommand{\Id}{\mathrm{I}}
\newcommand{\Ln}{\mathrm{L}}

\newcommand{\gH}{\gate{H}}

\newcommand{\C}{\mathbb{C}}

\newcommand{\myCirc}[1]{\mbox{\Qcircuit @C=1.em @R=1.2em {#1}\null\mbox{}} }
\newcommand{\myCircCentered}[1]{\begin{center}
\mbox{\Qcircuit @C=1.em @R=1.2em {#1}\null\mbox{}} 
\end{center}\vspace{1ex}}

\newcommandx{\circEq}[1][1=3ex]{\rule{0pt}{3ex}\multirow{2}*{=}}
\newcommandx{\circDots}{\dots}
\newcommand{\cNG}[1]{\gate{#1}}

\newcommand{\group}[1]{\mathbf{#1}}

\title{Constructive quantum scaling of unitary matrices}
\author[$*$,$\dagger$]{Adam Glos}
\author[$*$]{Przemysław Sadowski}

\affil[$*$]{Institute of Theoretical and Applied Informatics, Polish Academy
of Sciences, Ba{\l}tycka 5, 44-100 Gliwice, Poland}
\affil[$\dagger$]{Institute of Mathematics, Silesian University of Technology, 
Kaszubska 23, Gliwice 44-100, Poland}

\begin{document}

\maketitle

\begin{abstract}
In this work we present a method of decomposition of arbitrary unitary matrix
$U\in\group U(2^k)$ into a product of single-qubit negator and
controlled-$\sqrt{\mbox{NOT}}$ gates. Since the product results with negator
matrix, which can be treated as complex analogue if bistochastic matrix, our
method can be seen as complex analogue of Sinkhorn-Knopp algorithm, where
diagonal matrices are replaced by adding and removing an one-qubit
ancilla. The decomposition can be found constructively and resulting circuit
consists of $O(4^k)$ entangling gates, which is proved to be optimal. An example
of such transformation is presented.
\end{abstract}
 
 

\section{Introduction} 

Scaling a real matrix $O$ with non-negative entries means finding diagonal
matrices $D_1,D_2$  such that $B=D_1OD_2$ is bistochastic. Sinkhorn theorem
presents a necessary and sufficient condition for existence of the 
decomposition of a matrix. Moreover, the iterative Sinkhorn-Knopp algorithm 
finds
the bistochastic matrix $B$ \cite{sinkhorn1967concerning}. Such decomposition
can be used for ranking web pages \cite{knight2008sinkhorn}, preconditioning
sparse matrices \cite{livne2004scaling} and understanding traffic 
circulation~\cite{knight2012fast}.

Since unitary matrices are complex analogue of orthogonal matrices, it is
natural to ask whether there exist a counterpart of Sinkhorn theorem for them.
De Vos and De Baerdemacker considered whether it is possible, that for arbitrary
unitary matrix $U\in\group U(n)$ there exist two unitary diagonal matrices
$U_1,U_2$ such, that matrix $U_1UU_2$ has all lines sums equal to~1.  Such
decomposition exists for arbitrary unitary matrix and an algorithm for
finding it approximately was presented \cite{vos2014scaling}. Matrices called
\emph{negators} were treated as quantum counterpart of bistochastic matrices and
form a group $\group{XU}(n)$ under multiplication. Idel and Wolf propose an 
application of the quantum scaling in quantum optics \cite{idel2015sinkhorn}. 

Algorithm converges for arbitrary unitary matrix $U$ \cite{de2015two}. Similar
decomposition of unitary matrices $U\in\group{U}(2m)$ called $bZbXbZ$
decomposition was presented~\cite{fuhr2015biunimodular}. They show, that there
always exist matrices $A,B,C,D\in\group{U}(m)$ such that
\begin{equation}U = \begin{bmatrix}
A & 0 \\ 0 & B
\end{bmatrix}
\frac{1}{2}\begin{bmatrix}
\Id+C & \Id-C \\ \Id-C & \Id+C
\end{bmatrix}
\begin{bmatrix}
\Id & 0 \\ 0 & D
\end{bmatrix},
\end{equation}
where $\Id$ is identity matrix. Matrix in the middle is a block-negator matrix 
(which is also a negator matrix),
while left and right matrices are block diagonal matrices. In
\cite{de2015sinkhorn} an algorithm of finding such decomposition was presented.

Group $\group{XU}(2^n)$ is isomorphic to $\group{ U}(2^n-1)$ and can be
generated by single-qubit negator and controlled-$\sqrt{\mbox{NOT}}$ gates
\cite{de2013negator}. However, the proof is non-constructive since a
decomposition designed for generating random matrices was used
\cite{pozniak1998composed}. Although it is proved that it exists for any unitary
matrix, obtaining such a decomposition is a very complex task. Therefore another
approach is needed for efficient decomposition procedure.

In this article, using similar method to presented by de Vos and de Baerdemacker
\cite{de2013negator}, we demonstrate an implementation of arbitrary $k$-qubit
unitary operation using one-qubit ancilla with controlled-$ \sqrt{\textrm{NOT}}$
and single-qubit negator gates. Since product of these basic negator gates is
still a negator matrix, our result can be seen as quantum analogue of scaling
matrix. More precisely we prove, that for arbitrary matrix $U\in \group U(2^k)$,
which is performed on system $\Hl$, there exist a negator
$N\in\group{XU}(2^{k+1})$ such that for arbitrary state $\ket{\psi}\in \Hl$ we
have
\begin{equation}
U\ket{\psi} = \Psi(N \Phi(\ket{\psi})).
\end{equation}
Here $\Phi$ denotes the operation of extending the system with an ancilla 
register in $\ket{-}$
state and $\Psi$ denotes partial trace over the ancilla system. Since after 
performing operations $\Phi$ and $N$ the state is of the form $\ket{-}\kron 
U\ket{\psi}$, the partial trace is simply removing the ancilla system giving a 
pure state $U\ket{\psi}$. We 
describe an
efficient algorithm that for given $U$ returns explicit and exact form of $N$ 
with 
decomposition
into a sequence of  single-qubit negator and
controlled-$\sqrt{\textrm{NOT}}$ gates only in contrast to results of de Vos 
and de Baerdemacker \cite{de2015sinkhorn,de2013negator}.


In Section 2 we recall basic facts. In Section 3 we show how to perform such
transformation efficiently and demonstrate the cost in term of controlled-$
\sqrt{\textrm{NOT}}$ gates.
To illustrate the
transformation method, a transformation of Grover's search algorithm is
presented step by step in Section 4. 

\section{Basic facts}

Negator  gates of dimension 2 were introduced by de Vos and de Baerdemacker
\cite{de2013negator} as unitary matrices $N\in\group U(2)$ which are also a
convex combination of identity matrix and NOT gate. Simple calculation shows,
that they are of the form
$$N(\theta)=\frac{1}{2} \begin{bmatrix}
1 + e^{i\theta} & 1 - e^{i\theta} \\
1 - e^{i\theta} & 1 + e^{i\theta}
\end{bmatrix},$$
where $\theta\in[0,2\pi]$.
Negators form a subgroup of single-qubit unitary operations, i.e. $N(\phi)N(\psi)=N(\phi+\psi)$ for any values of $\phi$ and $\psi$. In the following we will also use a 2-qubit negator operation controlled-$\sqrt{\textrm{NOT}}$ gate (which is also controlled-$N(\frac{\pi}{2})$ gate) 
$$\begin{bmatrix}
1 & 0 & 0 & 0 \\
0 & 1 & 0 & 0 \\
0 & 0 & \frac{1+i}{2} & \frac{1-i}{2} \\
0 & 0 & \frac{1-i}{2} & \frac{1+i}{2} \\
\end{bmatrix}.$$
As these gates are used as basic operators, we will use a simplified notation in circuit, respectively
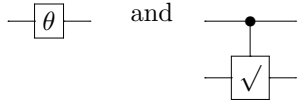
\begin{center}
\begin{tabular}{ccccccc}
\mbox{\myCirc {
& \cNG{\theta} & \qw \\
}}& and&
\mbox{\myCirc {
& \ctrl{1} & \qw \\
& \gate{\sqrt{}} & \qw }}
\end{tabular}.
\end{center} \vspace{1ex}
These two kinds of unitary matrices will be called NCN gates (\textit{Negators-Controlled-Negator}).

In Section 3 decomposition of single-qubit unitary gates will be needed. Every unitary matrix $U\in 
\group{U}(2)$ can
be presented as a product of global phase, two $z$-rotators and one $y$-rotator \cite{nielsen2010quantum}
\begin{equation}\label{eq:unitary2Dec}
\begin{split}
U &= e^{i\phi_0} 
\begin{bmatrix}
\cos \frac{\phi_1}{2} e^{i\phi_2} & \sin\frac{\phi_1}{2} e^{i\phi_3}\\
-\sin\frac{\phi_1}{2} e^{-i\phi_3} & \cos \frac{\phi_1}{2} e^{-i\phi_2}
\end{bmatrix}  \\
&=
e^{i\phi_0}
\begin{bmatrix}
e^{i\frac{\phi_2+\phi_3}{2}} & 0 \\
0 & e^{-i\frac{\phi_2+\phi_3}{2}}
\end{bmatrix}
\begin{bmatrix}
\cos \frac{\phi_1}{2} & \sin\frac{\phi_1}{2}\\
-\sin\frac{\phi_1}{2} & \cos \frac{\phi_1}{2} 
\end{bmatrix}
\begin{bmatrix}
e^{i\frac{\phi_2-\phi_3}{2}} & 0 \\
0 & e^{-i\frac{\phi_2-\phi_3}{2}}
\end{bmatrix}  \\
&= e^{i\phi_0} R_z(-\phi_2-\phi_3)R_y(\phi_1)R_z(\phi_3-\phi_2).
\end{split}
\end{equation}
Since global phase is not measurable, we can simplify this representation without loss of information 
\begin{equation} U \cong R_z(\gamma)R_y(\beta)R_z(\alpha), 
\label{eq:unitary2DecSimp}\end{equation}
where `$\cong$' means equality up to a global phase. The same applies in the case of global phase change on one of the registers of a bigger system
\begin{equation}
U_1\kron e^{i\phi} U_2 \kron U_3 = e^{i\phi}(U_1\kron U_2\kron U_3) \cong U_1\kron U_2\kron U_3 .
\end{equation}
Using these two facts we can say that in any situation we can ignore global 
phase change on any register.

 While it may lead to a conclusion that our 
transformation is mainly applied to group $\group{SU}(n)$, we decided to stay 
with the unitary matrices formalism, since negator gates are not special 
unitary matrices. The result may be written using the special matrices, however 
then the negators gates column and row sums will equal
$e^{i\theta}$ in general. 

\section{Circuit transformation method}

 In this section we provide complete description of the transformation method.
 We recall a sketch of a proof of universality theorem between quantum gates and
 negator gates from the work of de Vos and de Baerdemacker \cite{de2013negator}.
 Next we present transformation method of arbitrary single-qubit gate into NCN
 product. Then we provide a method of decomposition for arbitrary $k$-qubit
 circuit, based on the single qubit case. Finally, we analyse the cost of
 presented transformation.

 \subsection{Universality theorem}
 
De Vos and de Baerdemacker proved a universality theorem: group $\group
{XU}(2^k)$ generated by negators and controlled-$ \sqrt{\textrm{NOT}}$ is
isomorphic to $\group U(2^k-1)$ \cite{de2013negator}. The proof consists of
several steps:
\begin{enumerate}
\item Every matrix $U\in \group U(2^k-1)$ can be decomposed into a  product of 
$m$ gates $U_1U_2\dots U_m$, where matrices $U_i\in \group U(2^k-1)$ are of 
some special forms \cite{pozniak1998composed}.
\item Group $\group U(2^k-1)$ is isomorphic to group 
\begin{equation}\group {^{1}U}(2^k) = \left \{\begin{bmatrix} 1 & \mathbf 0 ^T 
\\ \mathbf 0 & U\end{bmatrix}: U\in \group U(2^k-1) \right \},
\end{equation}
because of the isomorphism $h: \group U(2^k-1) \to \group {^{1}U}(2^k) $
\begin{equation}
h(U) = \begin{bmatrix}
1 & \mathbf 0 \\
\mathbf 0 & U
\end{bmatrix}.
\end{equation}
\item  Function $f : \group {^{1}U}(2^k) \to \group {XU}(2^k)$ of the form 
$f(U)=(H\kron\Id_{2^k})U(H\kron\Id_{2^k})$ is an isomorphism.
\item Decomposition of every $f(h(U_i))$ into a product of NCN gates is 
possible, where $U_i$ comes from point 1.
\end{enumerate}
The proof used the decomposition presented in the work of Poźniak, Życzkowski and Kuś \cite{pozniak1998composed}, because it is proven that the decomposition exists for any unitary matrix. However obtaining such decomposition is a very complex task. Therefore we need to choose a different decomposition in order to find an efficient decomposition procedure.

Obviously, group $\group U(2^k)$ is isomorphic to some subgroup of
$\group{XU}(2^{k+1})$. In other words, with ancilla (one additional qubit) every
unitary matrix can be replaced with a sequence of NCN gates. For our purpose we
choose function $g:\group U(2^k) \to  \group{XU}(2^{k+1})$
\begin{equation}
g(U) = \frac{1}{2}H\kron \Id (\ketbra{0}{0}\kron \Id + 
\ketbra{1}{1}\kron U)H\kron \Id = \frac{1}{2}
\begin{bmatrix}
\Id & \Id \\ \Id & -\Id
\end{bmatrix}
\begin{bmatrix}
\Id & \mathbf 0\\ 
\mathbf 0 & U
\end{bmatrix}\begin{bmatrix}
\Id & \Id \\ \Id & -\Id
\end{bmatrix}. \label{eq:isomorphism}
\end{equation}
 Using the function $g$, every gate $U$ changes into controlled-$U$. Using 
 circuit 
 notation we can present this fact as
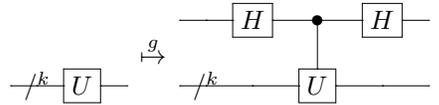
\myCircCentered {
&&&&\rule{0pt}{3ex}\multirow{2}*{$\overset{g}{\mapsto}$} && \qw & \gH &\ctrl{1} 
& \gH &\qw \\
&{/^k} \qw&\gate{U} & \qw & && {/^k} \qw& \qw & \gate{U} & \qw &\qw &  
\hspace{-1eM}.
}
Note that if we assume that the first qubit is set to $\ket{-}$, the control 
qubit does not influence the result (the condition is always `true'). 

\subsection{Single-qubit gate transformation}

Now we aim at decomposition of arbitrary single-qubit gate into NCN gates.
With Eq. (\ref{eq:unitary2DecSimp}) for any $U\in \Ln(\C^2)$ there exist real parameters $\alpha,\beta,\gamma$ such that
\begin{equation}U \cong R_z(\gamma)R_y(\beta)R_z(\alpha).\end{equation}
Therefore after applying function $g$ we have
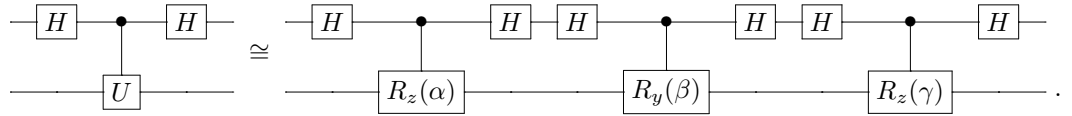
\myCircCentered {
& \gH & \ctrl{1} & \gH & \qw & \rule{0pt}{3ex}\multirow{2}*{$\cong$} && \gH & \ctrl{1} & \gH & \gH & \ctrl{1} & \gH & \gH & \ctrl{1} & \gH & \qw \\
& \qw & \gate{U} & \qw & \qw & && \qw & \gate{R_z(\alpha)} & \qw & \qw & 
\gate{R_y(\beta)} & \qw & \qw & \gate{R_z(\gamma)} & \qw & \qw & \hspace{-1eM}.
}
We change the rotators with neighbouring Hadamard gates into NCN gates as in
Fig.~(\ref{fig:basicGatesDecomposition})
\begin{center}\mbox{
\Qcircuit @C=.5em @R=1.2em{
& \gH & \ctrl{1} & \gH & \qw & \rule{0pt}{3ex}\multirow{2}*{$\cong$} && 
\cNG{\frac{\alpha}{2}} & \qw & \cNG{-\frac{\alpha}{2}} & \qw & \qw &
\ctrl{1} & \targ & \ctrl{1} &  \cNG{\frac{\beta}{2}} & \ctrl{1} & 
\cNG{-\frac{\beta}{2}} 
& \ctrl{1}  & \targ & \ctrl{1}&\qw &
\cNG{\frac{\gamma}{2}} & \qw & \cNG{-\frac{\gamma}{2}} & \qw & \qw \\
& \qw & \gate{U} & \qw & \qw & && 
\ctrl{-1} & \targ & \ctrl{-1} & \targ &  \qw &
\targ & \ctrl{-1}  &\gate{\sqrt{}^\dagger} & 
\qw& \targ & \qw& \gate{\sqrt{}^\dagger} & \ctrl{-1} & \targ & \qw &
\ctrl{-1} & \targ & \ctrl{-1} & \targ &  \qw & \hspace{-0.5eM}.
}
}\end{center}\vspace{1ex}
Let us note that the symbols of controlled-NOT,
controlled-$\sqrt{\textrm{NOT}}^\dagger$ and controlled-negator used in the
decomposed circuit do not mean that these gates cannot be transformed. We use
these symbols as a simplified notation for its decomposition with use of
controlled-$\sqrt{\textrm{NOT}}$ gates as shown in Fig.~(\ref{fig:NotsNot}).

\begin{figure}\centering
\mbox{\myCirc {
& \gH & \ctrl{1} & \gH & \qw & \circEq	&& 
\ctrl{1} & \targ & \ctrl{1} &  \cNG{\frac{\theta}{2}} & \ctrl{1} & 
\cNG{-\frac{\theta}{2}} 
& \ctrl{1}  & \targ & \ctrl{1}&\qw \\
& \qw & \gate{R_y(\theta)} & \qw & \qw & && 
\targ & \ctrl{-1}  &\gate{\sqrt{}^\dagger} & 
\qw& \targ & \qw& \gate{\sqrt{}^\dagger} & \ctrl{-1} & \targ & \qw
}}\\[1eM]
\mbox{\myCirc {
& \gH & \ctrl{1} & \gH & \qw & \circEq	&& 
\cNG{\frac{\theta}{2}} & \qw & \cNG{-\frac{\theta}{2}} & \qw & \qw \\
& \qw & \gate{R_z(\theta)} & \qw & \qw & && 
\ctrl{-1} & \targ & \ctrl{-1} & \targ &  \qw
}}\\[1eM]
\mbox{\Qcircuit @C=.5em @R=1.2em{
&\gate{H} & \ctrl{1} &\gate{H} & \qw & \circEq && \qw & \gate{-\frac{\pi}{2}} & 
\ctrl{1} & \gate{\frac{\pi}{2}} & \ctrl{1} & \qw & \gate{-\frac{\pi}{2}} & 
\ctrl{1} & \gate{\frac{\pi}{2}} & \ctrl{1} &\gate{\sqrt{}} & 
\gate{-\frac{\pi}{4}} & \ctrl{2} & \gate{-\frac{\pi}{4}} & \ctrl{2} & 
\gate{\frac{\pi}{2}} & \qw 
\\
&\qw & \ctrl{1} &\qw & \qw & && \ctrl{1} & \gate{-\frac{\pi}{2}} & 
\targ & \qw & \targ & \ctrl{1} & \gate{-\frac{\pi}{2}} & 
\targ & \qw & \targ & \qw & \qw & \qw & \qw & \qw & \qw & \qw \\
&\qw & \targ & \qw & \qw & && \gate{\sqrt{}} & \qw &  \qw & \qw & \qw 
& \gate{\sqrt{}^\dagger}& \qw &  \qw & \qw & \qw & \ctrl{-2} & 
\gate{-\frac{\pi}{4}} & \targ & \qw & \targ & \gate{\frac{\pi}{2}}\qw 
}\null\mbox{}}\vspace{1ex}
\caption{Decomposition of controlled-$y$-rotator, controlled-$z$-rotator and 
Toffoli gate. Decompositions use the simplified notation from 
Fig.~\ref{fig:NotsNot}.}
\label{fig:basicGatesDecomposition}
\end{figure}
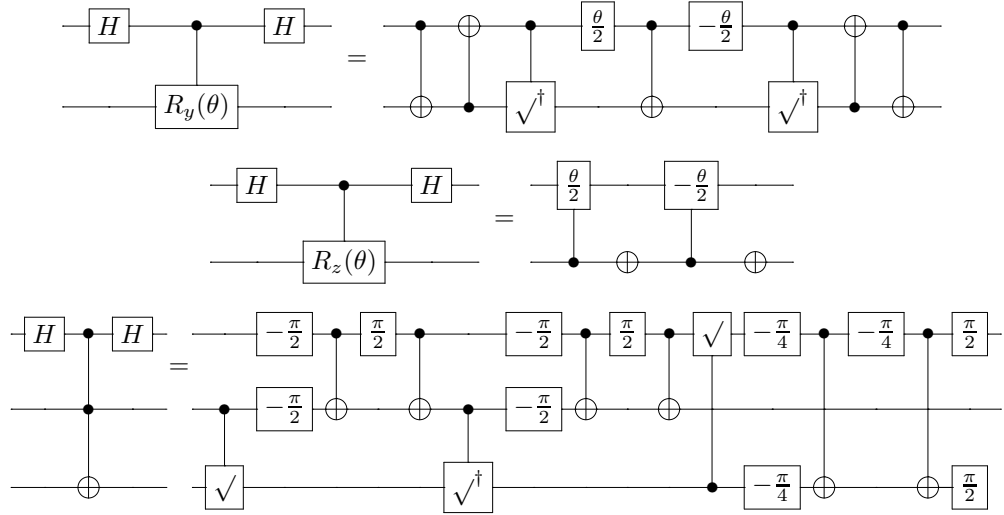
\begin{figure}
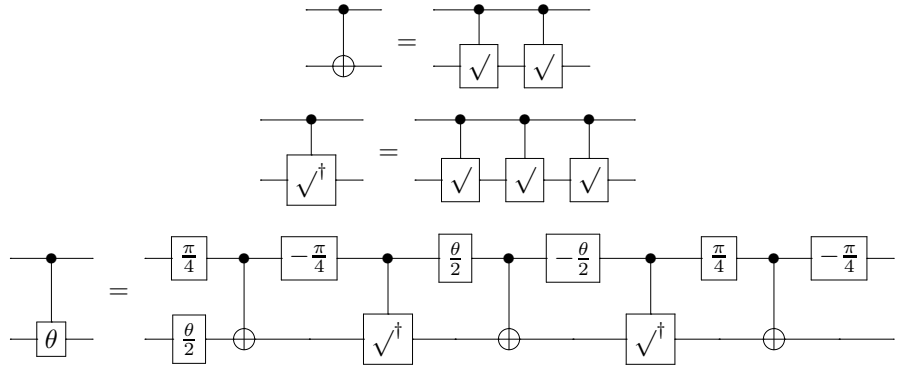
\centering
\mbox{\myCirc {
 &\ctrl{1} & \qw &\circEq && \ctrl{1} & \ctrl{1} &\qw \\
 &\targ & \qw &  && \gate{\sqrt{}}& \gate{\sqrt{}}& \qw
}}\\[1eM]
\mbox{\myCirc {
 &\ctrl{1} & \qw & \circEq && \ctrl{1} & \ctrl{1} & \ctrl{1} &\qw \\
 & \gate{\sqrt{}^\dagger} & \qw &  && \gate{\sqrt{}}& \gate{\sqrt{}}& 
\gate{\sqrt{}} & \qw
}}\\[1eM]
\mbox{ \myCirc {
& \ctrl{1} & \qw & \circEq && \gate{\frac{\pi}{4}} & \ctrl{1} &
\gate{-\frac{\pi}{4}} & \ctrl{1} & \gate{\frac{\theta}{2}} & \ctrl{1} & 
\gate{-\frac{\theta}{2}} & \ctrl{1} & \gate{\frac{\pi}{4}} & \ctrl{1} &
\gate{-\frac{\pi}{4}} & \qw \\
& \gate{\theta} & \qw & && \gate{\frac{\theta}{2}} & \targ & \qw & 
\gate{\sqrt{}^\dagger} & \qw & \targ &  \qw & \gate{\sqrt{}^\dagger} & \qw & 
\targ & \qw & \qw 
}}
\caption{Decomposition of controlled-NOT, controlled-$ 
\sqrt{\textrm{NOT}}^\dagger$ gates and controlled-negator~\cite{de2013negator}.}
\label{fig:NotsNot}
\end{figure}

\subsection{General transformation method}
Now we consider transformation of arbitrary $k$-qubit circuit.
Let us assume that we have a circuit which consists of unitary operations 
$U\in\Ln(\C^{2^k})$, generalized measurement $\mathbf M = 
\{M_a\in\Ln(\C^{2^k}):a\in\Sigma\}$, where $\Sigma$ is a set of classical 
outputs of measurement, and starting state $\ket{\phi_0}$
\myCircCentered {
\lstick{\ket{\phi_0}}&{/^k} \qw& \gate{U} & \measure{\mathbf{M}} & \hspace{-1eM}.
}
In order to construct a decomposition of unitary $U$ into a sequence of negator 
gates we 
begin with obtaining a decomposition of $U$ into controlled-NOT and 
single-qubit gates 
\myCircCentered {
\lstick{\ket{\phi_0}}&{/^k} \qw& \gate{U} & \measure{\mathbf{M}} & \push\cong & 
\push{\ket{\phi_0}} &{/^k} \qw& \gate{V_1} & \gate{V_2} &\qw& \cdots && 
\gate{V_m} & \measure{\mathbf{M}} & \hspace{-1eM},
}
here denoted by a sequance of gates $U=V_m\cdots V_1$. 
Contrary to the decomposition 
presented in the work of Poźniak, Życzkowski and Kuś, there exist efficient 
methods for constructing such circuit~\cite{mottonen_quantum_2004}.
Next we need to add an additional qubit, transform $V_i$ gates into controlled-$V_i$ gates and add Hadamard gates as below (since $HH=\Id$)
\myCircCentered {
\lstick{\ket{1}} & \qw&\gH & \gH & \ctrl{1} & \gH & \gH & \ctrl{1} & \gH &\qw & 
\cdots && \gH & \ctrl{1} & \gH &\gH  &\qw &\ket{1} \\ 
\lstick{\ket{\phi_0}} &{/^k} \qw& \qw & \qw & \gate{V_1} &\qw &\qw & \gate{V_2} 
\qw & \qw & \qw& \cdots && \qw & \gate{V_m} &\qw & \qw & \measure{\mathbf{M}} 
& \hspace{-1eM}.
}
Let us note that product $H\cdot \textrm{controlled-}V_j\cdot H$ is an image of 
homomorphism presented in Eq.~(\ref{eq:isomorphism}) on $V_j$. 
Next we replace the product with the sequence of NCN gates (here denoted by $\mathbf N_j$) as in previous subsection (if $V_j$ is  controlled-NOT, then we choose Toffoli gate transformation from Fig. (\ref{fig:basicGatesDecomposition}))
\myCircCentered{
\lstick{\ket{1}} & \qw &\gH & \multigate{1}{\mathbf{N_1}} & 
\multigate{1}{\mathbf{N_2}} &\qw & \cdots && \multigate{1}{\mathbf{N_m}} &\gH & \qw &
\ket{1} \\
\lstick{\ket{\phi_0}} & {/^k} \qw&\qw & \ghost{\mathbf{N_1}} & 
\ghost{\mathbf{N_2}} & \qw &\cdots && \ghost{\mathbf{N_m}} &\qw & 
\measure{\mathbf{M}} & \hspace{-1eM}.
}
For the sake of simplicity we may change the starting state and resulting state 
on the first wire
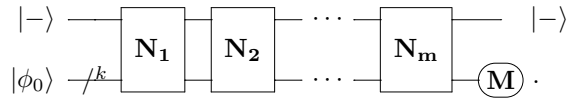
\myCircCentered {
\lstick{\ket{-}} & \qw &\multigate{1}{\mathbf{N_1}} & \multigate{1}{\mathbf{N_2}} &\qw & \cdots && \multigate{1}{\mathbf{N_m}} & \qw & \ket{-}  \\
\lstick{\ket{\phi_0}} &{/^k} \qw& \ghost{\mathbf{N_1}} & \ghost{\mathbf{N_2}} 
&\qw & \cdots && \ghost{\mathbf{N_m}} & \measure{\mathbf{M}} & \hspace{-1eM}.
}

Now we have an equivalent circuit which consists of negators and controlled-$ \sqrt{\textrm{NOT}}$ gates only.

\subsection{Transformation cost}
Now we consider upper bound of cost of decomposition into negator circuit. Two kinds will be discussed: memory complexity and number of single and two-qubit gates. In the first case for arbitrary $k$-qubit circuit transformation requires one additional qubit.

Let $c_{\textrm{CNOT}}(k) $ and $c_s(k)$ denote upper bound of the number of 
respectively controlled-NOT and single qubit-gates needed for the 
implementation of an arbitrary $k$-qubit circuit. Using the operation presented 
above we need $17c_{\textrm{CNOT}}(k) + 64c_s(k)$ controlled-$ 
\sqrt{\textrm{NOT}}$ gates and $11c_{\textrm{CNOT}}(k)+34c_s(k)$ negators to 
implement an equivalent circuit (up to global phase).

Any circuit which consists of controlled-NOT and single-qubit gates can be 
simplified in such a way, that $c_s(k)\leq 2c_{\textrm{CNOT}}(k)+k$. This 
estimation is based on the worst case, when there are two single-qubit gates 
between every controlled-NOT gate. Taking this into account we 
can express the previous result in terms of $c_{\textrm{CNOT}}$ only, because 
only $17c_{\textrm{CNOT}}(k)+ 64c_s(k) \leq 145 c_{\textrm{CNOT}}(k)+64k$ 
controlled-$ \sqrt{\textrm{NOT}}$ gates are needed. In fact, if 
$c_{\textrm{CNOT}} = O(4^k)$, then so is the number of controlled-$ 
\sqrt{\textrm{NOT}}$ gates.

\begin{figure}[t]
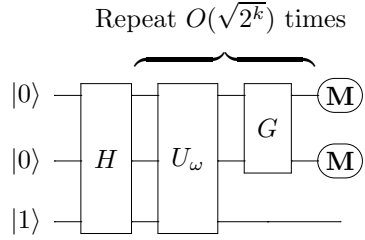

\hspace{1eM}
\begin{center}\mbox{
\myCirc {
& & \ustick{\text{\hspace{2.5eM}Repeat $O(\sqrt{2^k})$ times}}\\
\lstick{\ket{0}} & \multigate{2}{H} & \multigate{2}{U_\omega} & \multigate{1}{G} & \measure{\textbf{M}} \\
\lstick{\ket{0}} & \ghost{H} & \ghost{U_\omega} &  \ghost{G}  & \measure{\textbf{M}} \\
\lstick{\ket{1}} & \ghost{H} & \ghost{U_\omega} & \qw & \qw 
\gategroup{2}{3}{2}{4}{1.7eM}{^\}}
 }
}\end{center}
\caption{Original Grover's search algorithm circuit in case $k=2$. $G$ is 
Grover diffusion 
operator, $U_\omega$ is quantum black box and we perform measurement $M$. 
Algorithm 
comes from~\cite{lavor2008grover}.}
\label{fig:originalGroverAlgorithm}
\end{figure}

\newcommand{\circGroverLongGrover}{
\makebox[10cm]{\raisebox{-0.5cm}[0cm]{
\myCirc {
&  &  &&&&&  \ustick{\text{\hspace*{-2eM}Repeat $O(\sqrt{2^k})$ 
times}}\\
\lstick{\ket{1}} & \gate{H} & \gate{H} &\ctrl{1} &  \gate{H} &  \qw 
& \gate{H} & \ctrl{1} & \gate{H} & \gate{H} & \qw & \qw &\rstick{\ket{1}} \\
\lstick{\ket{0}}  & \qw& \qw &\multigate{2}{H}&  \qw & 
\multigate{2}{U_\omega} & \qw & \multigate{1}{G}   & \qw & 
\qw &  \measure{\textbf{M}} & \cw \\
\lstick{\ket{0}}   & \qw & \qw& \ghost{H} &  \qw &  \ghost{U_\omega} & 
 \qw & \ghost{G}         & \qw & \qw & \measure{\textbf{M}} & \cw\\
\lstick{\ket{1}}   & \qw & \qw& \ghost{H} & \qw &  
\ghost{U_\omega}  	   &\qw&\qw&\qw&\qw & \qw & \qw       
\gategroup{2}{6}{2}{9}{1.7eM}{^\}}
 }}
}
}
\newcommand{\circGroverbegin}{
\makebox[1.8cm]{\raisebox{-0.2cm}[0.2cm][1.5cm]{
\myCirc {
\lstick{\ket{-}} & \qw & \qw \\
\lstick{\ket{0}} & \qw & \qw \\
\lstick{\ket{0}} & \qw & \qw \\
\lstick{\ket{1}} & \qw & \qw \\
} }}
}
\newcommand{\circGroverEnd}{
\raisebox{-0.2cm}[0.2cm][2.2cm]{
\myCirc {
 &\qw&  \ket{-} & \\
 &\qw& \measure{\textbf{M}} & \cw\\
&\qw & \measure{\textbf{M}} & \cw\\
 &\qw& \qw & \qw\\
}}
}
\newcommand{\circGroverHadamardStretched}{
\myCirc{
&\gate{H} & \ctrl{1} & \gate{H} & \gate{H} & \ctrl{2} & \gate{H} & \gate{H} & \ctrl{3} & \gate{H} &  \qw  \\
&\qw & \gate{H} & \qw & \qw & \qw & \qw & \qw & \qw & \qw &\qw   \\
&\qw & \qw & \qw &\qw & \gate{H} & \qw & \qw & \qw & \qw & \qw  \\
&\qw & \qw & \qw &\qw & \qw & \qw &\qw & \gate{H} & \qw &\qw 
}
}
\newcommand{\circGroverHadamardDecomposed}{
\mbox{
\Qcircuit @C=.5em @R=1.2em{
& \cNG{\frac{\pi}{2}} & \qw & \cNG{\frac{-\pi}{2}} & \qw & \qw &
\ctrl{1} & \targ & \ctrl{1} &  \cNG{\frac{\pi}{4}} & \ctrl{1} & 
\cNG{-\frac{\pi}{4}} & \ctrl{1}  & \targ & \ctrl{1}&\qw & \qw \\
& \ctrl{-1} & \targ & \ctrl{-1} & \targ &  \qw &
\targ & \ctrl{-1}  &\gate{\sqrt{}^\dagger} & 
\qw& \targ & \qw& \gate{\sqrt{}^\dagger} & \ctrl{-1} & \targ & \qw & 
\qw
}
}
}

\newcommand{\circGroverCnotStretched}{ 
\Qcircuit @C=.5em @R=1.2em{ &\gate{H} &
\ctrl{1} & \gate{H} & \gate{H} & \ctrl{2} & \gate{H} & \qw  \\
 &\qw & \targ & \qw & \qw & \qw & \qw & \qw \\ &\qw & \qw & \qw &\qw & \targ & 
 \qw & \qw 
 } }
 
\newcommand{\circGroverGroverDiffusionOperator}{ 
\Qcircuit @C=.5em @R=1.2em{ 
&\gate{H} & \ctrl{1} &\gate{H} &\gate{H} & \ctrl{1} & \gate{H} & \gate{H} &
\ctrl{1} & \gate{H} & \gate{H} & \ctrl{2} & \gate{H} & \gate{H} & \ctrl{1} &
\gate{H} & \gate{H} & \ctrl{2} & \gate{H} & \gate{H} & \ctrl{1} & \gate{H} &
\gate{H} & \ctrl{1} & \gate{H} & \qw\\
 & \qw & \multigate{1}{H}& \qw & \qw & \targ\qwx[1] & \qw & \qw & \ctrl{1} & 
 \qw & \qw & \qw & \qw & \qw & \ctrl{1} & \qw & \qw   \qw & \qw & \qw & \qw & 
 \targ\qwx[1] & \qw & \qw & \multigate{1}{H} & \qw & \qw \\ 
 & \qw & \ghost{H} & \qw & \qw & \targ & \qw & \qw & \targ & \qw &
\qw & \gate{R_z\left (- \frac{\pi}{2} \right )} & \qw & \qw &\targ & \qw & \qw &
\gate{R_z\left ( \frac{\pi}{2}\right )} & \qw & \qw &  \targ & \qw & \qw &
\ghost{H} & \qw &  \qw 
} }

\newcommand{\circGroverToffoliDecomposed}{
\mbox{\Qcircuit @C=.5em @R=1.2em{
& \qw & \gate{-\frac{\pi}{2}} & 
\ctrl{1} & \gate{\frac{\pi}{2}} & \ctrl{1} & \qw & \gate{-\frac{\pi}{2}} & 
\ctrl{1} & \gate{\frac{\pi}{2}} & \ctrl{1} &\gate{\sqrt{}} & 
\gate{-\frac{\pi}{4}} & \ctrl{2} & \gate{-\frac{\pi}{4}} & \ctrl{2} & 
\gate{\frac{\pi}{2}} & \qw 
\\
& \ctrl{1} & \gate{-\frac{\pi}{2}} & 
\targ & \qw & \targ & \ctrl{1} & \gate{-\frac{\pi}{2}} & 
\targ & \qw & \targ & \qw & \qw & \qw & \qw & \qw & \qw & \qw \\
& \gate{\sqrt{}} & \qw &  \qw & \qw & \qw 
& \gate{\sqrt{}^\dagger}& \qw &  \qw & \qw & \qw & \ctrl{-2} & 
\gate{-\frac{\pi}{4}} & \targ & \qw & \targ & \gate{\frac{\pi}{2}}\qw 
}\null\mbox{}}
}

\newcommand{\circGroverRotatorDecomposed}{
\Qcircuit @C=.5em @R=1.2em{
& \cNG{-\frac{\pi}{4}} & \qw & \cNG{\frac{\pi}{4}} & \qw & \qw \\
& \ctrl{-1} & \targ & \ctrl{-1} & \targ &  \qw
}}

\newcommand{\circGroverCnotDecomposed}{
\Qcircuit @C=.8em @R=1.2em{
& \cNG{\frac{\pi}{2}} & \qw & \cNG{\frac{-\pi}{2}} & \qw & \qw &
\ctrl{1} & \targ & \ctrl{1} &  \cNG{\frac{\pi}{2}} & \ctrl{1} & 
\cNG{-\frac{\pi}{2}} & \ctrl{1}  & \targ & \ctrl{1}&\qw & \qw \\
& \ctrl{-1} & \targ & \ctrl{-1} & \targ &  \qw &
\targ & \ctrl{-1}  &\gate{\sqrt{}^\dagger} & 
\qw& \targ & \qw& \gate{\sqrt{}^\dagger} & \ctrl{-1} & \targ & \qw & 
\qw
}
}

\begin{figure}
 \footnotesize

\centering

\begin{tikzpicture}[scale=0.85,transform shape]

\definecolor{colBeg}{rgb}{1,0.,0.};
\definecolor{colHadamard}{rgb}{0.,.5,0.};
\definecolor{colGrover}{rgb}{0.3,0.3,1};
\definecolor{colEnd}{rgb}{1,0.,1};

\tikzstyle{begBranchColor} = [draw=colBeg]
\tikzstyle{hadamardBranchColor} = [draw=colHadamard]
\tikzstyle{groverBranchColor} = [draw=colGrover]
\tikzstyle{endBranchColor} = [draw=colEnd]

\tikzstyle{nodeStyle} = [anchor=north west ,draw, thick]
\tikzstyle{nodeLastStyle} = []
\tikzstyle{squareRound} = [dashed, thick ]; 
  
\tikzset{
  myarrow/.style={
    decoration={markings,mark=at position 1 with {\arrow[#1]{stealth}}},
    postaction={decorate},
    shorten >=0.4pt,
    squareRound,thick,
    draw=#1},
  myarrow/.default=blue}

\node (LGn) at (0,0) {};
\node (Bn) at (-1.5,-4.2) {};
\node (HSn) at (1,-4.2) {};
\node (En) at (9.2,-4.2) {};
\node (HDn) at (0.64,-7.6) {};
\node (GDOn) at (-2.5,-9.8) {};
\node (TOFn) at (1.5,-12.4) {};
\node (CSn) at ( -2.5,-12.4) {};
\node (RDn) at (11.5 ,-12.4) {};
\node (CDn) at (-2.5 ,-15.4) {};

\node (LG) at (LGn)[nodeStyle]{\circGroverLongGrover};

\node (B)  at (Bn) [nodeStyle,begBranchColor] {\circGroverbegin};
\node (HS) at (HSn) [nodeStyle,hadamardBranchColor] {\circGroverHadamardStretched};
\node (E) at (En) [nodeStyle,endBranchColor]{\circGroverEnd};

\node (HD) at (HDn)[nodeStyle,hadamardBranchColor]  {\circGroverHadamardDecomposed};

\node (GDO) at (GDOn) [nodeStyle,groverBranchColor] {\circGroverGroverDiffusionOperator};

\node (TOF) at (TOFn)[nodeStyle,groverBranchColor]  
{\circGroverToffoliDecomposed};

\node (RD) at (RDn)  [nodeStyle,groverBranchColor]{\circGroverRotatorDecomposed};
\node (CS) at (CSn) [nodeStyle,groverBranchColor] {\circGroverCnotStretched};
\node (CD) at (CDn) [nodeStyle,groverBranchColor] {\circGroverCnotDecomposed};

\draw [squareRound,begBranchColor] ($(LGn)+(0.2,-0.85)$) rectangle 
($(LGn)+(1.62,-3.55)$);
\draw [squareRound,hadamardBranchColor] ($(LGn)+(1.72,-0.85)$) rectangle 
($(LGn)+(4.,-3.55)$);
\draw [squareRound,groverBranchColor] ($(LGn)+(5.05,-0.85)$) rectangle 
($(LGn)+(7.3,-2.9)$);
\draw [squareRound,endBranchColor] ($(LGn)+(7.4,-0.85)$) rectangle 
($(LGn)+(9.9,-3.55)$);

\draw [squareRound,hadamardBranchColor] ($(HSn)+(0.32,-0.06)$) rectangle ($(HSn)+(2.4,-1.28)$);

\draw [squareRound,groverBranchColor] ($(GDOn)+(2.1,-0.06)$) rectangle 
($(GDOn)+(3.65,-2.05)$);
\draw [squareRound,groverBranchColor] ($(GDOn)+(3.71,-0.06)$) rectangle 
($(GDOn)+(5.24,-2.05)$);
\draw [squareRound,groverBranchColor] ($(GDOn)+(5.29,-0.06)$) rectangle 
($(GDOn)+(7.95,-2.05)$);

\draw [squareRound,groverBranchColor] ($(CSn)+(0.2,-0.06)$) rectangle 
($(CSn)+(1.75,-1.28)$);

\node (Bna) at  ($(LGn)+(0.3,-3.45)$) {};
\node (HSna) at ($(LGn)+(2.86,-3.45)$) {};
\node (HDna) at ($(HSn)+(1.36,-1.28)$) {};
\node (Ena) at ($(LGn)+(9.8,-3.55)$) {};
\node (GDOna1) at ($(LGn)+(6.175,-2.9)$) {};
\node (GDOna2) at ($(GDOna1)+(2.5,-1)$) {};
\node (TOFna) at ($(GDOn)+(4.475,-2.05)$) {};
\node (RDna1) at ($(GDOn)+(6.62,-2.05) $) {};
\node (RDna2) at ($(GDOn)+(15.3,-2.25) $) {};
\node (CDna) at ($(CSn) +(0.975,-1.28)$) {};
\node (CSna) at ($(GDOn)+(2.875,-2.05)$) {};

\draw[myarrow=colBeg] (Bna) -- node[right] {a)} (Bna|-Bn) ;
\draw[myarrow=colHadamard] (HSna) -- node[right] {c)} (HSna|-HSn);
\draw[myarrow=colHadamard] (HDna) -- node[label={[xshift=0.3cm, yshift=-0.95cm]d)}] {} (HDna|-HDn);
\draw[myarrow=colEnd] (Ena) -- node[right] {b)}  (Ena|-En);
\draw[myarrow=colGrover] (GDOna1) -- (GDOna1|-GDOna2) -- ($(GDOna2)+(0,0)$) -- 
node[right] {c)}  (GDOna2|-GDOn);
\draw[myarrow=colGrover] (TOFna) -- node[right] {d)}  (TOFna|-TOFn);
\draw[myarrow=colGrover] (CDna) --node[right] {d)}  (CDna|-CDn);
\draw[myarrow=colGrover] (CSna) -- node[right] {c)}  (CSna|-CSn);
\draw[myarrow=colGrover] (RDna1) -- (RDna1|-RDna2) -- ($(RDna2)+(0,0)$) -- 
node[right] {d)}  (RDna2|-RDn);




\end{tikzpicture}\vspace{1eM}
\caption{Grover's search algorithm decomposition. Unnecessary Hadamard gates 
have already been removed a) $\ket{1}$ changes into $\ket{-}$; b) measurement 
$\mathbf{M}$ does not change, on first wire we end with $\ket{-}$ state; c) 
subcircuit simplification; d) subcircuit transforming into NCN gates. Any other 
transformation left can be done similarly, except the $U_\omega$ case.} 
\label{fig:GroverDecomposition}
\end{figure}
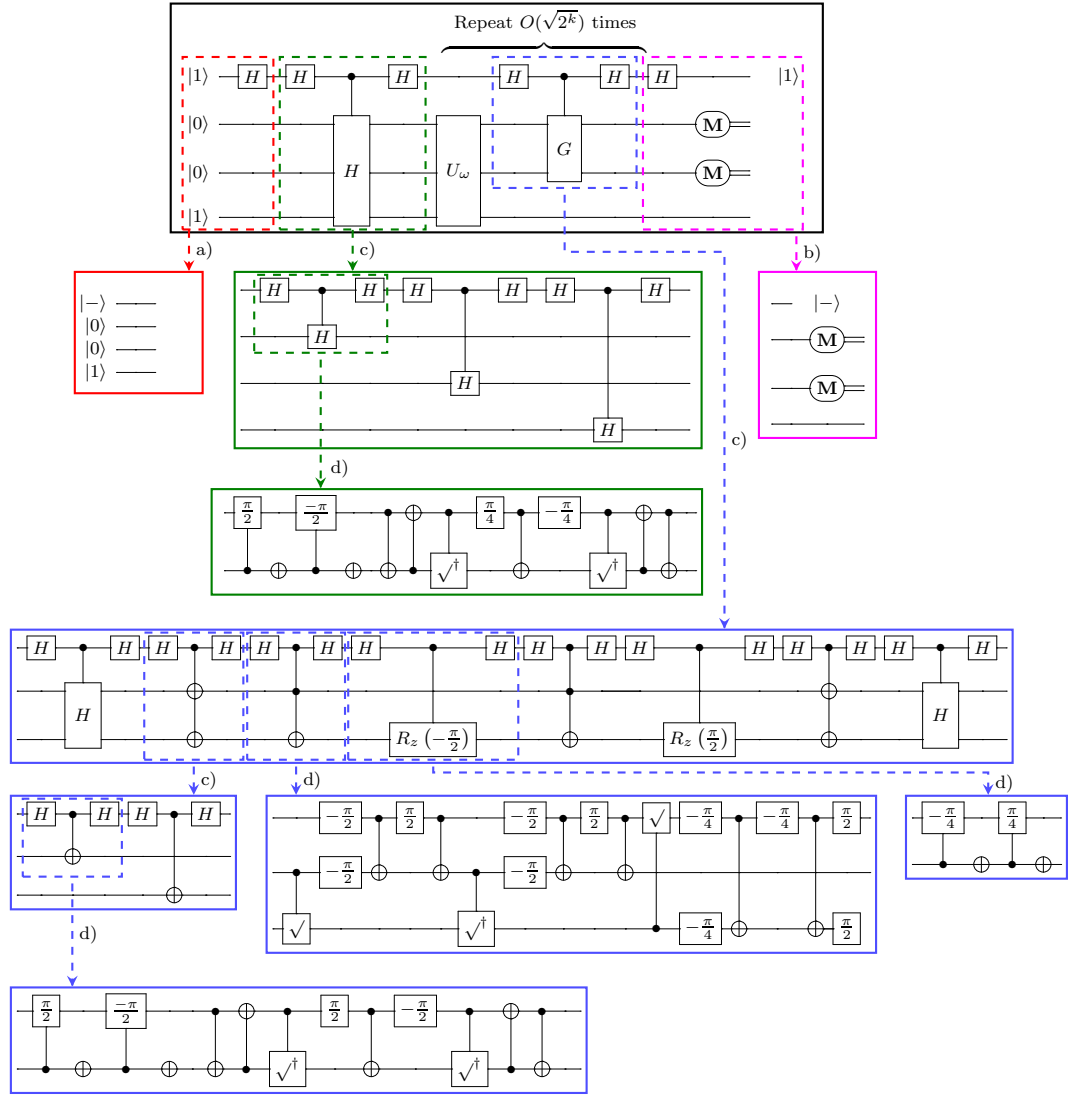

\section{Step by step transformation example}
To illustrate the introduced decomposition we will present Grover's algorithm 
for $k=2$ qubits as NCN circuit. The original circuit for this algorithm is 
presented in Fig. (\ref{fig:originalGroverAlgorithm}), where $\omega$ denotes 
the searched state.

As in the previous section, we will add one qubit, change every $H$ and $G$ 
gate into controlled-$H$ and controlled-$G$ respectively and add Hadamard gates 
on the ancilla register. Former steps of the decomposition are explicitly 
presented in Fig.~(\ref{fig:GroverDecomposition}). The following facts were used
\begin{itemize}
\item the decomposition of Hadamard gate is $H\cong 
R_z(\pi)R_y(\frac{\pi}{2})R_z(0)=R_z(\pi)R_y(\frac{\pi}{2})$, 
\item the decomposition of NOT gate is $\mbox{NOT}\cong 
R_z(\pi)R_y(\pi)R_z(0)=R_z(\pi)R_y(\pi)$,
\item for any $U,V\in\Ln(\C^2)$ we have
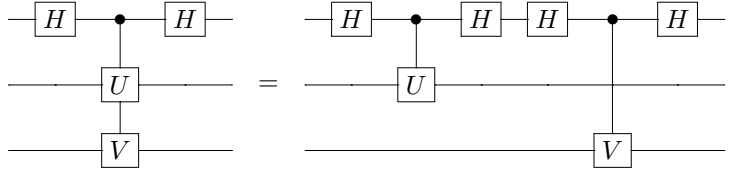
\begin{center}\mbox{
\myCirc{
& \gate{H} & \ctrl{1} & \gate{H} & \qw &&& \gate{H} & \ctrl{1} & \gate{H} & 
\gate{H} & \ctrl{2} & \gate{H} & \qw \\
& \qw & \gate{U} \qwx[1] & \qw & \qw & \push{=}&& \qw & \gate{U} &\qw &\qw & 
\qw & \qw &\qw \\
& \qw & \gate{V} & \qw & \qw & && \qw & \qw & \qw &\qw &\gate{V} &\qw & \qw & 
\hspace{-1eM},\\
}
}\end{center}
\item Grover's diffusion operator can be decomposed in the following way
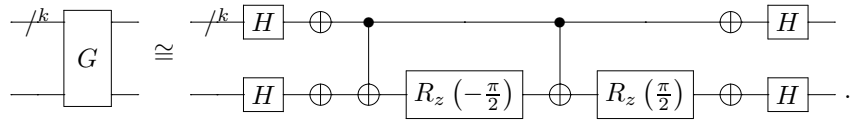
\begin{center}\mbox{
\myCirc {
& {/^k}\qw & \multigate{1}{G} & \qw & \rule{0pt}{3ex}\multirow{1}*{$\cong$} && 
{/^k}\qw & \gate{H} & \targ & \ctrl{1} & \qw & \ctrl{1} & \qw & \targ & 
\gate{H} & \qw \\
& \qw & \ghost{G} & \qw & && \qw & \gate{H} & \targ & \targ & \gate{R_z\left 
(-\frac{\pi}{2}\right )} & \targ & \gate{R_z\left (\frac{\pi}{2}\right )} 
&\targ 
& \gate{H} & \qw & \hspace{-1eM}.\\
}}\end{center}
\end{itemize}

Decomposition of $U_\omega$ depends strictly on the value of $\omega$, 
therefore it is not
presented in the example. The full decomposition is presented in Fig.
(\ref{fig:GroverDecomposition}).

\section{Concluding remarks}
In the presented work we provide a constructive method of scaling arbitrary 
unitary matrices $U\in \group U(2^k)$. More precisely we proved that for 
arbitrary unitary matrix $U\in \group U(2^k)$ there exists unitary negator 
matrix $N\in \group{XU}(2^{k+1})$ such that for arbitrary state $\ket{\psi}$ we 
have
\begin{equation}
U\ket{\psi} = \Psi(N \Phi(\ket{\psi})).
\end{equation}
Here $\Phi$ denotes the operation of extending the system with an ancilla
register in $\ket{-}$ state and $\Psi$ denotes partial trace over the ancilla
system. We described efficient algorithm of decomposing $N$ into product of
single-qubit negator and controlled-$\sqrt{\mbox{NOT}}$ gates. Our decomposition
consists of $O(4^k)$ entangling gates which is proved to be optimal and needs
one qubit ancilla.

Our result can be seen as complex analogue of Sinkhorn-Knopp algorithm, which is
known to have wide applications. The result is in contrast to the previous
results \cite{de2013negator}, which could be only used to prove the existence of
such decomposition. Moreover, our transformation is exact and can be found 
constructively. In contrast to \cite{de2015sinkhorn}, our transformation 
consists only of negator gates. The main difference is that transformation 
needs one-qubit ancilla.

\section*{Acknowledgements}
The work was supported by the Polish National Science Centre: A. Glos under the research project number DEC-2011/03/D/ST6/00413, P. Sadowski under the research project number 2013/11/N/ST6/03030.

\bibliographystyle{ieeetr}
\bibliography{negatorBib}

\end{document}